\def\beq{\begin{equation}}
\def\eeq{\end{equation}}
\def\bmulteq{\begin{eqnarray}}
\def\emulteq{\end{eqnarray}}

\def\Tr{{\rm Tr}}

\def\>{\rangle}
\def\<{\langle}
\def\logtwo{{\rm log_2}}

\documentstyle[aps,preprint]{revtex}

\begin{document}

\preprint{
\noindent
\begin{minipage}[t]{3in}
\begin{flushright}
quant-ph/9808007 \\
\end{flushright}
\end{minipage}
}

\title{Entanglement of projection and a new class of quantum erasers}

\author{Robert Garisto}
\address{BNL Theory Group, Bldg 510a, Brookhaven National Lab, Upton, NY 11973}

\author{Lucien Hardy}
\address{Clarendon Laboratory, Oxford OX1 3PU, England}

\maketitle

\begin{abstract}
We define a new measurement of entanglement, the entanglement of
projection, and find that it is natural to write the entanglements of
formation and assistance in terms of it.  Our measure allows us to
describe a new class of quantum erasers which restore entanglement
rather than just interference. Such erasers can be implemented with
simple quantum computer components.  We propose realistic
optical versions of these erasers.

\ 

\noindent
PACS: 03.65.Bz,03.67.Lx
\end{abstract}

%
Entanglement is the degree to which the wave function does not
factorize.  For example, an $S=0$ two particle system $|+$$-$$\> -
|-$$+$$\>$ is maximally entangled: measurement of the spins reveals
they are completely anticorrelated.  
The concept of entanglement goes
to the very heart of quantum mechanics, and understanding its nature
is a prerequisite to understanding quantum mechanics itself.
Two-particle entanglement was used by Einstein, Podolsky and Rosen
\cite{EPR} to argue that quantum mechanics could not be a complete 
description of reality---that there had to be an underlying local
theory.  But J.~S.~Bell used such entangled states to show that any
local underlying theory would have to satisfy certain inequalities,
which quantum mechanics explicitly violates \cite{Bell}.  Experiments
on such entangled states have shown that these inequalities are
violated just as quantum mechanics predicts \cite{Aspect}.  Modern
research on entanglement includes proposals for providing
cleaner demonstrations of this nonlocality using three-particle
entangled states \cite{GHZ}, and on quantifying 
entanglement \cite{Bennett,vedral,wootters,entang_a}.

The goal of this Letter is to define a new class of quantum erasers
which restore entanglement of a multistate subsystem, rather than just
interference, and to quantify that restoration with a new measure of
entanglement \cite{phystoday}.  A quantum eraser \cite{scully} is a
device in which coherence appears to be lost in a subset of
the system, but in which that coherence can be restored by erasing the
tagging information which originally ``destroyed'' it.

Traditional erasers \cite{theo erasers,expt erasers} need only two
distinct subsystems.  For example, if one sends particle $A$ through
two slits, and if one ``tags'' which slit $A$ goes through via the
interaction with a tagging particle, $T$, then the interference
pattern will disappear.  But if one makes the ``which slit''
information in $T$ unobservable, even in principle, then one can
restore the interference pattern for $A$. To avoid the use of a
double-negative, one could refer to this as an {\it interference
restorer}.

A simple way to erase this tagging information is to measure $T$ in
the $|0\>_T \pm |1\>_T$ basis.  Here ``$|0\>_T $'' means ``T
interacted with $A$ at slit 0''.  The positions of $A$ on the screen
corresponding to $T$ in the state $|0\>_T +|1\>_T $ display an
interference pattern, and those corresponding to $T$ in the state
$|0\>_T -|1\>_T$ display a shifted interference pattern.  While the
overall pattern on the screen shows no interference, for the subsets
of these events corresponding to $|0\>+|1\>$ or $|0\>-|1\>$, coherence
is restored.

Our new class of erasers involves at least three subsystems, $A$, $B$
and $T$.  Consider an entangled state $|00\>_{AB} + |11\>_{AB}$ of
subsystem $AB$.  If we tag the pieces of this with $T$ so that the
wave function of the whole system is $|00\>_{AB}|0\>_T +
|11\>_{AB}|1\>_T$, then the entanglement of subsystem $AB$ appears
to be lost.  But if one erases that tagging information,
then the entanglement is restored.  Thus we will refer to this object
as a {\it disentanglement eraser}, or, equivalently, as an {\it
entanglement restorer}.


In order to discuss these new erasers, we will need to define several
measures of entanglement.  For a {\it pure}, two-particle, two-state
system which can be thought of as a pair of qubits (quantum bits), the
entanglement is well-defined.  One can always write such a pure
``$2\times 2$'' system in the Schmidt basis so that $|\psi_{AB}\> =
\alpha |00\>_{AB} + \beta |11\>_{AB}$, with $\alpha$ and $\beta$
positive and real, and $\alpha^2+\beta^2=1$.  The $AB$ system has a
pure density matrix $\rho_{AB}= |\psi_{AB}\>\< \psi_{AB}|$, while the
subsystem for $A$ alone has a mixed density matrix $\rho_A= \Tr_B
\rho_{AB}$.  Then one can write entanglement of $AB$ in terms of the
quantum relative entropy \cite{Bennett},

\beq
E(\psi_{AB}) = - \Tr \left[ \rho_A \logtwo \rho_A \right] = e(\alpha^2),
\label{eq:E of psi}
\eeq

\noindent
where $e(x) = -[ x \logtwo x + (1-x) \logtwo (1-x) ]$. Since
$e_{min}=e(0)=e(1)=0$ and $e_{max}=e(1/2)=1$, $E(\psi)$ ranges from 0,
for no entanglement, to 1, for a fully entangled state. This $E(\psi)$
remains constant with any unitary operation on $A$ or $B$, and is
changed only by operations where the effect on one state (say $A$)
depends upon another (either $B$ or a third state).  Such interactions
can be implemented with controlled-NOT gates (c-NOTs). 
One can show that all logic gates of a quantum
computer can be constructed solely in terms of unitary operations on
individual qubits and on c-NOTs between the qubits.  Thus one role of
multiqubit logic gates is to change the entanglement between pairs of
qubits.

Mixed states, on the other hand, do not have a unique measure of
entanglement.  One reason is that the entanglement for a mixed state
depends upon the basis chosen for mixed density matrix $\rho_{AB}$.
Let us write $\rho_{AB}$ in terms pure states 
$|\chi_i\>_{AB}$, with weights $p_i$,

\beq 
\rho_{AB} = \sum_{i=0}^{m-1} p_i | \chi_i\> \< \chi_i |.
\label{eq:rhoAB chi}
\eeq

\noindent 
Note that the $\chi_i$'s need not be orthogonal. 
Here $m \geq n$, with $n$ being the number of nonzero eigenvalues of
$\rho_{AB}$.  If one naively tries to define the entanglement of
Eq. (\ref{eq:rhoAB chi}) as $\sum_i p_i E(\chi_i)$, one finds that it
depends on the $\chi_i$'s chosen \cite{entang_a}.  Instead
what has been done traditionally is to write the {\it entanglement of
formation} \cite{wootters},

\beq 
E_f(\rho_{AB}) = {\rm min} \sum_{i=0}^{m-1} p_i E(\chi_i),
\label{eq:Ef}
\eeq

\noindent
which is the {\it minimum} value of the naive measure over all
decompositions of $\rho_{AB}$.  Note that the simplest decomposition
which gives the minimum average entanglement has the same number of
pure states as there are nonzero eigenvalues of $\rho_{AB}$ ({\it
i.e.} $m_f=n$) \cite{wootters}.  Recently, a new measure has been
derived called the {\it entanglement of assistance}
\cite{entang_a}, $E_a$, which is just the {\it maximum} value of the
naive measure over all decompositions. 

To see what the basis dependence of the naive measure really means, let us
write the mixed state $\rho_{AB}$ in terms of a higher dimensional pure state,

\beq
| \Psi_{ABT} \> = \sum_{i=0}^{d_T-1} \sqrt{p_i} | \psi_i\>_{AB} | i\>_T,
\label{eq:PsiABT}
\eeq

\noindent
where $| i \>_T$ are $d_T$ orthonormal pure states of a set of
``tagging particles'' or ``taggants''.  If we trace over the $d_T$
tagging states of the pure density matrix $|\Psi_{ABT} \> \<
\Psi_{ABT} |$, we obtain $\rho_{AB}$ of Eq. (\ref{eq:rhoAB chi}) with
$d_T$ component pure states: $\{ |\chi_i\>\} = \{|\psi_i\>\}$ with
$m=d_T$.  This basis depends upon the chosen taggant basis,
$\{|i\>_T\}$.  For a {\it given} taggant basis, the entanglement is
well defined:

\beq
E_{p\{|i\>_T \}} (\Psi_{ABT}) = \sum_{i=0}^{d_T-1} p_i E(\psi_i),
\eeq

\noindent
where the entanglement of the component pure states, $E(\psi_i)$, is
given by Eq. (\ref{eq:E of psi}).
We call this the {\it entanglement of projection}
because it corresponds to the projection of the full pure state
$\Psi_{ABT}$ onto a given taggant basis to yield a mixed subsystem
$AB$ with an entanglement $E_p$.  What this means practically is that
if subsystem $AB$ is entangled with a taggant $T$, and one measures the
taggant in basis $\{|i\>_T\}$ , the resulting projected pure states of
$AB$ have an average entanglement equal to $E_{p\{|i\>_T\}}$.

If one measures the taggant in a different basis $| i' \>_T=U | i
\>_T$, then the entanglement of projection becomes the weighted
average $\sum_i p_i'E(\psi_i')$, where we have rewritten Eq. (\ref{eq:PsiABT})
in the new taggant basis: $| \Psi_{ABT} \> = \sum_{i=0}^{d_T-1}
\sqrt{p_i'} | \psi_i'\>_{AB} | i'\>_T$.  This shows that for a given pure state
$| \Psi_{ABT} \>$, $E_p$ takes on different values for different
choices of taggant basis---there is no unique measure of entanglement
for a mixed subsystem $AB$.  In fact, by taking the minimum and
maximum values of $E_{p\{U\}}$ over all possible taggant bases $U |
i\>_T$, one recovers the entanglements of formation and assistance,

\beq
E_f = {\rm min}_U E_{p\{U\}}, E_a = {\rm max}_U E_{p\{U\}},
\label{eq:EfEa}
\eeq

\noindent
and $E_p$ is bounded by $E_f$ and $E_a$: $E_f \leq E_{p\{U\}} \leq
E_a$.

Our formula for $E_f$ in Eq. (\ref{eq:EfEa}) is identical to that of
Eq. (\ref{eq:Ef}) because $d_T$ is always greater than or equal to the
number of pure states $m_f$ needed in the minimal decomposition of the
$2\times 2$ subsystem (since $m_f=n\leq d_T$).  However, it turns out
that there are cases where $m_a > n$ \cite{thapliyal}.  This means
that our $E_a$ depends on $d_T$, and thus can be smaller than the
$E_a$ of Ref. \cite{entang_a}.  Our $E_a$ measures the amount of
assistance a ``friend'' $T$ can give to $AB$ for a specific pure state
$\Psi_{ABT}$, whereas their $E_a$ measures how much assistance an
arbitrary $T$ leading to $\rho_{AB}$ could give.

To quantify the entanglement in our erasers, we need to take into
account whether or not the taggant has been measured.  Let us define
$h$ to be the number of outcomes resulting from any measurements of
$T$, and $P_j$ as the projection operator for outcome $j$, which
occurs with probability $q_j$ and results in $AB$ state $\rho_j$.  Then
we can define the {\it entanglement of projections' formation},

\beq
E_{pf} = \sum_{j=0}^{h-1} q_j E_f(\rho_j).
\eeq

\noindent
If no measurement has been performed on $T$, then $h=1$,
$\rho_0=\rho_{AB}$ and $E_{pf}=E_f$.  If a nondegenerate measurement is
performed on $T$, then $h=d_T$, the $\rho_j$ are all pure, and $E_{pf}$ is
just $E_p$ for the basis of $T$ defined by the projectors $\{P_j\}$.
For $E_{pf}$ to increase after a measurement, there must have been
entanglement between $T$ and $AB$.      

To illustrate the utility of $E_p$ and $E_{pf}$, consider a
pure system $ABT$ whose $2\times 2$ subsystem $AB$ is a mixed state of
only two pure states:
$|\Psi_{ABT} \> = \alpha |00\>_{AB} |0\>_T + \beta |11\>_{AB} |1\>_T$.
It is clear that $E_p=0$ in the taggant basis $\{|0\>_T, |1\>_T\}$,
and thus $E_f=0$ for subsystem $AB$.  Since $T$ has not been measured,
$E_{pf}=E_f=0$.  
But if we project the taggant onto basis $|i'\>_T=U | i\>_T$, with

\beq
U = \pmatrix{a & b \cr 
             -b^* & a^* \cr},
\eeq

\noindent
the entanglement of projection of $AB$ in that basis is

\beq
E_{p\{U\}} = p_0 e(a^2\alpha^2/p_0) + p_1 e(b^2 \alpha^2/p_1),
\label{eq:Ep 2e}
\eeq

\noindent 
with the probability of the taggant being projected into state $|0'\>$
being $p_0=a^2 \alpha^2 + b^2 \beta^2$, and with $p_1=1-p_0$. 
(For our choice of basis we can take $a$ and $b$ to be real.)
Note that, after some algebra,
Eq. (\ref{eq:Ep 2e}) can be rewritten as
$E_{p\{U\}} = e(\alpha^2) + e(a^2) - e(p_0)$.
For $\alpha^2=\beta^2=1/2$, $AB$ is in mixed state $\rho_{AB}= 1/2(
|00\>\<00| + |11\>\<11|)$ and the entanglement of projection depends
on the choice of projection basis: $E_p=0$ ($E_p=1$) for $a^2=0$
($a^2=1/2$).  Thus $E_f=0$, $E_a=1$, and $E_{pf}$ is between $0$ to
$1$, depending on which basis one uses to measure $T$.  For
$\alpha^2=1$, $AB$ is in the pure state $\rho_{AB}= |00\>\<00|$, and
$E_p=0$ in all bases, so that $E_a=E_f=E_{pf}=0$.

Before we use these definitions on our new erasers, we need to briefly
address the entanglement of a $2\times 4$ subsystem---where subsystem
$B$ has dimension 4 instead of 2.  If the ``$B$'' part of $\Psi_{ABT}$
can be written just using $|0\>_B$ and $|1\>_B$, and not $|2\>_B$ and
$|3\>_B$, then the $AB$ subsystem can simply be treated like the
$2\times2$ case above.  On the other hand, if $\Psi_{ABT}$ can be
written in the form

\bmulteq
|\Psi_{ABT}\> &=& {1 \over 2} 
\bigl\{ \left[ |00\>_{AB}+ |11\>_{AB}\right] |0\>_T \nonumber\\
&+& \left[ |02\>_{AB} + |13\>_{AB}\right] |1\>_T \bigr\},
\label{eq:Psi 2x4}
\emulteq

\noindent
then no rotation of the taggant basis will change the entanglement of
the two component pure states, and thus the $AB$ mixed state is
unambiguously fully entangled ($E_{pf}=E_f=E_p=E_a=1$). 


Disentanglement erasers can be divided into two kinds: reversible and
irreversible.  Reversible erasers restore entanglement by simply
undoing the tagging operation which caused the apparent
disentanglement.  Consider Fig 1a, which starts with a fully entangled
pure state

\beq
|\Psi_{ABT}\> = {1 \over \sqrt{2}} \left[ | 00\>_{AB} + |11\>_{AB} \right]
|0\>_T.
\label{eq: Psi before}
\eeq

\noindent
By design, $E_{pf}=E_f=E_a=1$.  Now let $A$ (or $B$) act as the
controller in a c-NOT on $T$ in what we call the {\it tagger} (or,
alternatively, the {\it diluter}, since it dilutes the entanglement of
$AB$ into the full $ABT$ state \cite{Bennett}).  This puts $ABT$ into
GHZ state

\beq
|\Psi_{ABT}\> = {1 \over \sqrt{2}} \left[ | 000\>_{ABT} + |111\>_{ABT} 
\right],
\label{eq: Psi after}
\eeq

\noindent
whose entanglement of projection's formation is zero: $E_{pf}=E_f=0$.
Note that $E_a$ is still 1, which is the best possible $E_{pf}$
achievable after erasure. 

We accomplish the erasure in Fig. 1a simply  by passing $ABT$ through
the same c-NOT.  This {\it untagger} acts as a {\it concentrator} of
entanglement into $AB$ \cite{Bennett}.  The wave function is left in
the state of Eq. (\ref{eq: Psi before}) with $E_{pf}=E_f=1$.
Entanglement has thus been restored.

On the other hand, the eraser of Fig. 1b is irreversible.  The
entangled state of Eq. (\ref{eq: Psi before}) again goes through a
tagger, producing the state of Eq. (\ref{eq: Psi after}) with
$E_{pf}=E_f=0$.  But now we erase the tagging information by
measuring the taggant in some basis.  Unlike the reversible eraser,
this can be done as a delayed choice ({\it i.e.},after the measurement
of $A$ and $B$).  If $T$ is measured in basis $\{U\}$ defined above,
then $E_{pf}$ is just given by $E_{p\{U\}}$ in Eq. (\ref{eq:Ep 2e}).
In particular, if one measures $T$ in basis $|0\>_T \pm |1\>_T$ (so
that $a^2=1/2$), then $E_{pf}=E_p=1$, and thus entanglement is fully
restored.


Two optical experiments in Fig. 2 illustrate the workings of
entanglement restorers.  Both use two-photons states produced from a
parametric down conversion crystal, and both are feasible with current
technology.  But an entanglement restorer needs three separate states,
so we need to use more than one quantum number on each
particle---namely their spin and path ($i.e.$, position)
\cite{zukowski,rome}. In regions where no spin-path interactions occur, the
states cannot interact, even though they are on the same photon.  So
the states behave as if they were spatially separated.

The reversible eraser in Fig. 2a uses the two photon spins as the $AB$
subsystem and one of their paths as the taggant.  We can write the initial
wave function as

\beq
|\Psi^{rev.}_{t=0}\> = {1 \over\sqrt{2}} 
 \left[ |hv\>_{s_1s_2}-|vh\>_{s_1s_2}\right] | 0\>_{p_1},
\label{eq:psirev t0}
\eeq

\noindent
which can be written as (\ref{eq: Psi before}). 
Thus the $s_1$-$s_2$ subsystem of
Eq. (\ref{eq:psirev t0}) has $E_{pf}=E_f=1$.  By passing photon 1
through a polarizing beam splitter (PBS), we create a spin-path
interaction which is equivalent to a c-NOT on its path, giving

\beq
|\Psi^{rev.}_{t=1}\> = {1 \over\sqrt{2}} 
 \left[ |hv\>_{s_1s_2}|0\>_{p_1}-|vh\>_{s_1s_2} |1\>_{p_1}\right],
\label{eq:psirev t1}
\eeq

\noindent
which is the same as the tagged state in Eq. (\ref{eq: Psi
after}). Thus $E_{pf}=E_f=0$. What this means is that if one were to
measure the spins of the photons at this point (summing over paths $0$
and $1$), one would obtain a mixed state with $\rho=(|hv\>\< hv| +
|vh\>\< vh|)/2$, which could just as well have been formed from states
that were never entangled.  And while Eq. (\ref{eq:psirev t1}) is
technically a GHZ state, one cannot use it to perform an unambiguous
test of nonlocality because there are only two distinct locations for
the three states \cite{zukowski}.  
Still, any local effect mimicking GHZ correlations
would involve some novel spin-position interaction and so an
experimental test seems worthwhile.

To reversibly erase the tagging information at $t=2$, we simply
perform the reverse of the operation of $t=1$.  This PBS evolves
$\Psi$ back to Eq. (\ref{eq:psirev t0}), and thus entanglement is
restored: $E_{pf}=E_f=1$.  
Note that we could have instead constructed
an irreversible $s_1$-$s_2$ eraser by removing the second PBS and
measuring $p1$ in the $|0\>\pm|1\>$ basis.  But this could not be done
as a ``delayed choice'' since $s_1$ and $p_1$ are properties of the
same photon.

The irreversible eraser in Fig. 2b treats the spin and path of photon
1 as its subsystem $AB$, and the spin of the other photon as the
taggant.  This allows us to restore the entanglement of $AB$ $after$
the properties of $A$ and $B$ have been measured.  Since we start out
with the wave function of Eq. (\ref{eq:psirev t0}), we need to create
spin-path entanglement via $s_1$-$p_1$ interactions.  First we pass
photon 1 through a PBS oriented in the $h/v$ direction to obtain

\beq
|\Psi^{irrev.}_{t=1}\> = {1 \over\sqrt{2}}
 \left[ |h0\>_{s_1p_1}|v\>_{s_2}-|v2\>_{s_1p_1}|h\>_{s_2} \right].
\label{eq:psiirev t1}
\eeq

\noindent
This can be written as the tagged state $[|00\>_{AB}|0\>_T +
|12\>_{AB}|1\>_T]/\sqrt{2}$ which is of the same form as Eq. (\ref{eq:
Psi after}).  Here we have made $B$ a two-qubit subsystem since it
encompasses four separate paths.  So the operation at $t=1$ is a c-NOT
on the first qubit of $B$ by $A$.  The reason we put $ABT$ in a tagged
state first is that the only $AB$-$T$ interaction which takes place in
this eraser occurs in the down-conversion crystal at $t=0$. Thus we
need to preserve the taggant connection to $AB$ even in the fully
entangled state.

To create the $s_1$-$p_1$ entangled state, we pass photon 1 through a
pair of PBS's in the $\bar h/\bar v = (h+v)/(-h+v)$ direction, which
act as a c-NOT on the second qubit of $B$ by $A$ (in the $\bar 0/\bar
1$ basis):

\bmulteq
|\Psi^{irrev.}_{t=2}\> &=& {1 \over 2}
\left\{ \bigl[ |\bar h0\>_{s_1p_1}- |\bar v1\>_{s_1p_1}\right] |v\>_{s2}
\nonumber \\
&-& \left[ |\bar h2\>_{s_1p_1}+ |\bar v3\>_{s_1p_1}\bigr] |h\>_{s2}
\right\}.
\label{eq:psiirev t2}
\emulteq   

\noindent
This can be written as the $2\times4$ system in Eq. (\ref{eq:Psi 2x4}).
As we stated before, no rotation of $T$ for such a $2\times4$ system
will change $E_p$ from $1$, so that $E_{pf}=E_f=1$---the $s_1$-$p_1$
subsystem is fully entangled despite its connection to $s_2$.

To make $E_{pf}=0$, we simply reverse the last step to obtain the
tagged state of Eq. (\ref{eq:psiirev t1}) again.  Finally, we erase
the tagging information by measuring the taggant $s_2$ in some basis.
If we measure $s_2$ in the $h/v$ basis, the $AB$ subsystem is left in
a mixed state $\rho_{AB}=(|00\>\<00| + |12\>\<12|)/2$, and
$E_{pf}=E_p=0$.  But if we measure $s_2$ in the $\bar h/\bar v$ basis,
$AB$ is left in the mixed state $\rho_{AB}=([|00\> + |12\>][\<00|+
\<12|] + [|00\> -|12\>][\<00| -\<12|])/4$, whose component pure states
each are fully entangled.  Thus $E_{pf}=E_p=1$ and we can restore
$s_1$-$p_1$ entanglement even after photon 1 has been measured.


The entanglement of projection provides a new framework for
quantifying the entanglement of mixed states by thinking of them as
higher dimensional pure states.  It allows us to describe a new class
of quantum erasers, called entanglement restorers, which can be
thought of as simple quantum computer components.  They show how c-NOT
operations can shift entanglement from one part of the computer to
another.  It is possible that understanding how entanglement changes
in a quantum computer will aid in pinpointing the source of their
exponential speedup over classical computers.

Recently there has been considerable progress in manipulating three
and four photon states \cite{innsbruck}, although as of yet it has not
been possible to implement a c-NOT on two photons.  Once this
technological hurdle has been cleared, it will be possible to
construct three-particle disentanglement erasers.  Until that time,
the two-photon experiments described above should be able to test most
of their interesting features.

We thank David DiVincenzo, Ashish Thapliyal, and Tony Leggett for useful
comments.


\begin{figure}
\caption[]{
{\small (a) Entangled state $AB$ enters the tagger, which dilutes the
entanglement into the whole $ABT$ system. Reversing this operation
restores $AB$ entanglement. (b) After tagging, entanglement is
restored by measuring $T$.  }}
\end{figure}
\begin{figure}
\caption[]{
{\small
(a) At $t=1$, the $s_1$-$s_2$ entanglement of the two photons is
tagged by path $p_1$ via a PBS (c-NOT).  As in
Fig. 1a, the tagging operation is simply reversed. (b) The $s_1$-$p_1$
entangled state of $t=2$ is (re)tagged with $s_2$ at $t=3$.  Next one
measures $s_1$ and $p_1$ in a basis determined by the possible path BS
and orientation of the $s_1$ PBS analyzers.  Finally, as in Fig 1b,
one can restore entanglement by measuring $s_2$.
}}
\end{figure}

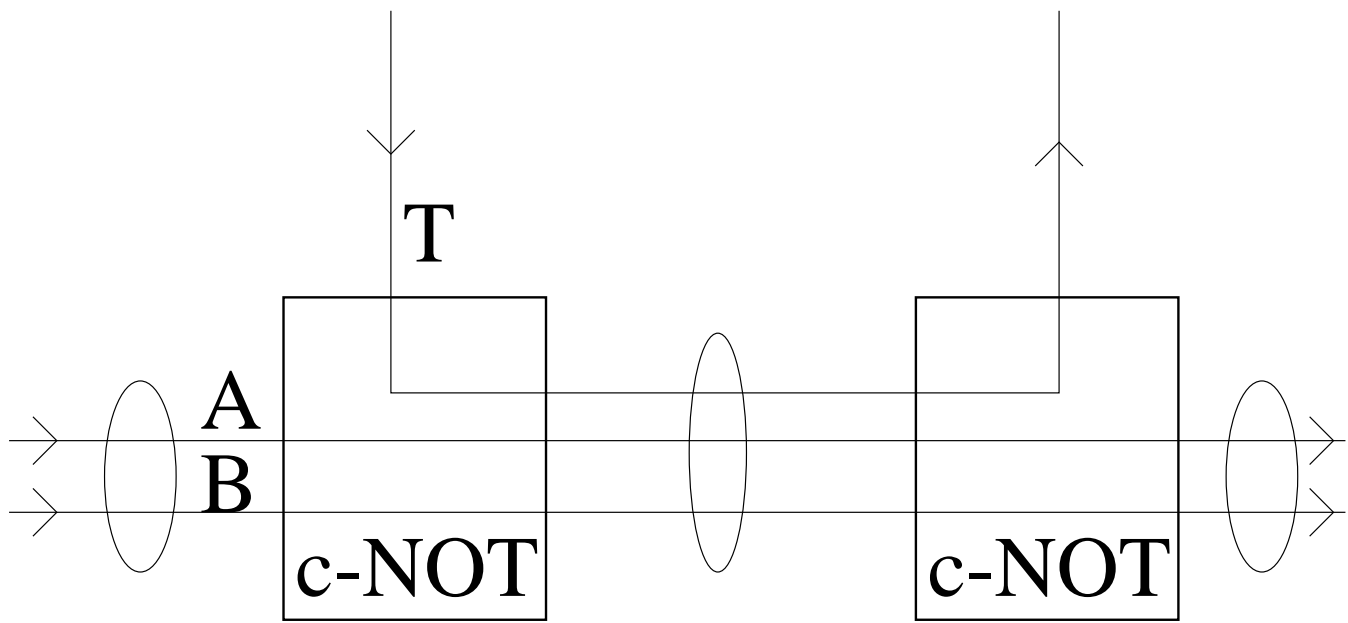
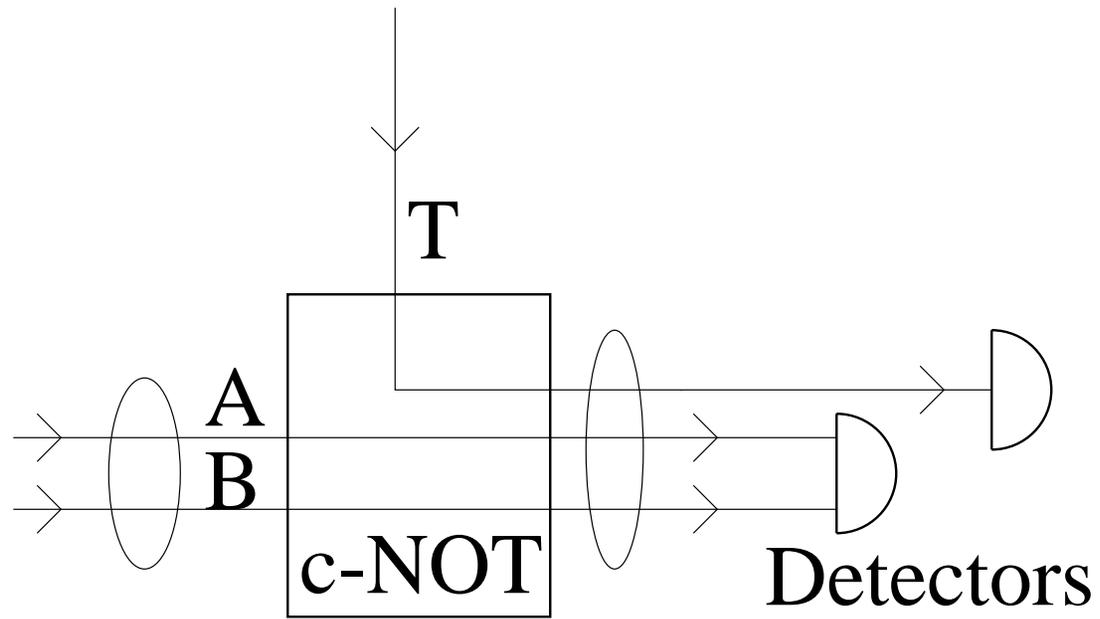
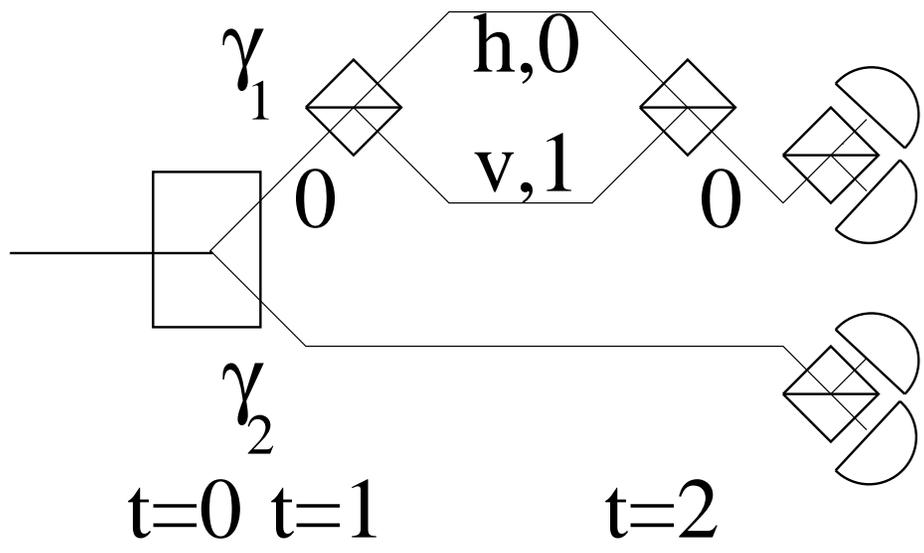
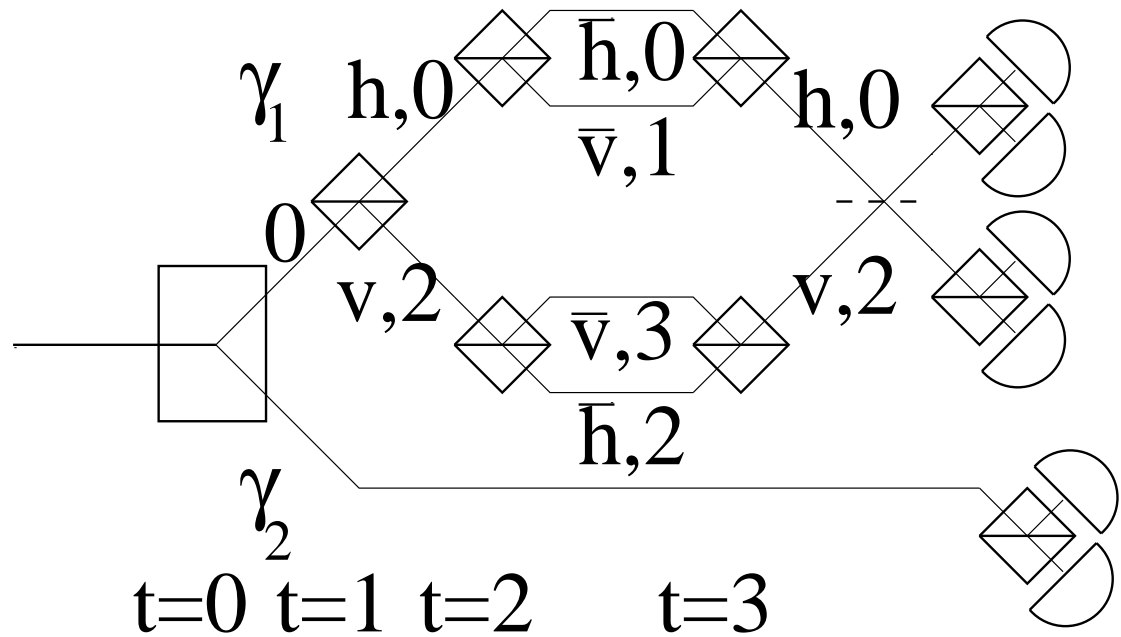
\end{document}